\documentclass[showpacs,amsmath,preprintnumbers,12pt]{revtex4}

\usepackage{amssymb}
\usepackage{graphicx}

\begin{document}

\title{Scattering of Elastic Waves in a Quasi-one-dimensional Cavity: Theory and Experiment}

\author{G. B\'aez} 
\affiliation{\'Area de F\'isica Te\'orica y Materia Condensada, Universidad
Aut\'onoma Metropolitana-Azcapotzalco, A. P. 21-267, 04000 M\'exico D.
F., Mexico}

\author{M. Cobi\'an-Su\'arez} 
\affiliation{\'Area de F\'isica Te\'orica y Materia Condensada, Universidad
Aut\'onoma Metropolitana-Azcapotzalco, A. P. 21-267, 04000 M\'exico
D. F., Mexico}

\author{A. M. Mart\'inez-Arg\"uello} 
\affiliation{Departamento de F\'{\i}sica, Universidad Aut\'onoma
Metropolitana-Iztapalapa, A. P. 55-534, 09340 M\'exico D. F., Mexico}

\author{M. Mart\'{\i}nez-Mares}
\affiliation{Departamento de F\'{\i}sica, Universidad Aut\'onoma
Metropolitana-Iztapalapa, A. P. 55-534, 09340 M\'exico D. F., Mexico}

\author{R. A. M\'endez-S\'anchez}
\affiliation{Instituto de Ciencias F\'isicas, Universidad Nacional Aut\'onoma de
M\'exico, A. P. 48-3, 62210 Cuernavaca Mor., Mexico}

\begin{abstract}
We study the scattering of torsional waves through a quasi-one-dimensional
cavity both, from the experimental and theoretical points of view. The
experiment consists of an elastic rod with square cross section. In order to
form a cavity, a notch at a certain distance of one end of the rod was grooved.
To absorb the waves, at the other side of the rod, a wedge, covered by an
absorbing foam, was machined. In the theoretical description, the scattering
matrix $S$ of the torsional waves was obtained. The distribution of $S$ is
given by Poisson's kernel. The theoretical predictions show an excellent
agreement with the experimental results. This experiment corresponds, in quantum
mechanics, to the scattering by a delta potential, in one dimension, located at
a certain distance from an impenetrable wall.
\end{abstract}


\maketitle

\section{Introduction}
The scattering of waves by cavities is a problem of interest in several areas of
physics. This is due to the fact that cavities present the majority of phenomena
observed in the scattering by complex
systems~\cite{MelloKumar,Beenakker,Alhassid}. The theoretical and numerical
studies of scattering by cavities are extensive including the simplest
one-dimensional ones~\cite{MelloKumar,Gopar,
Martinez-ArguelloMendez-SanchezMartinez-Mares, Dominguez-RochaMartinez-Mares}
and the two-dimensional cavities both, integrable and
chaotic~\cite{MelloBaranger,MelloMartinez-Mares,
Martinez-MaresBaezMendez-Sanchez,MelloGoparMendez-Bermudez,
Mendez-BermudezLuna-AcostaIzrailev}. The quantum graphs~\cite{KottosSmilansky}
which also display complex behavior can be considered as one-dimensional
scattering cavities. 

Up to now, scattering experiments by cavities have been performed using
mesoscopic cavities~\cite{Marcus}, quantum corrals~\cite{Crommie}, microwave
cavities~\cite{DoronSmilanskyFrenkel,LewenkopfMullerDoron,
Mendez-SanchezKuhlBarthLewenkopfStockmann,
SchanzeStockmannMartinez-MaresLewenkopf,
KuhlMartinez-MaresMendez-SanchezStockmann,Anlage,Mortessagne2006,Richter},
optical microcavities~\cite{NoeckelStone,Capassoetal} and microwave
graphs~\cite{Sirko}. In all these experiments the measurements are 
done in the frequency domain. Wave transport experiments on elastic systems, on
the other hand, are scarce and mainly performed in the time
domain~\cite{Fink,EPL2011}. In this paper we introduce a system in which the
transport of elastic waves can be studied in the frequency domain from both,
the theoretical and experimental points of view. 

We organize the paper as follows. In the next section we propose a theoretical
model for the scattering of torsional waves by a one-dimensional cavity in
an elastic beam. This is done by grooving of a rectangular notch in a specific
place of a semi-infinite beam. The scattering matrix $S$ of this system is
obtained and we show that its distribution is correctly described by Poisson's
kernel. In Sec.~\ref{sec:ExperimentalSetup} we describe the beam used in the
experiment: a notch in one side of the beam and a passive vibration isolation
system, on the other side. This beam allows the measurement of the scattering of
waves by the cavity formed by the notch and the free-end of the beam. In the
same section the experimental setup, used to measure the resonances of the
elastic cavity, is also presented. In Sec.~\ref{sec:Comparison} we compare the
analytical results with the experiment. Some brief conclusions are given in
Sec.~\ref{sec:Conclusions}.


\section{The Theoretical Model and Poisson's kernel}
\label{sec:TheoreticalModel}

\begin{figure}
\includegraphics[width=8.3cm]{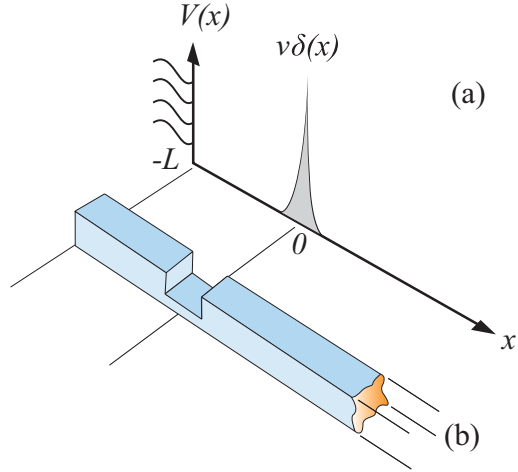}
\caption{(a) Quantum scattering cavity formed by a delta potential and an
impenetrable barrier. (b) An elastic scattering cavity formed by a notch on a
semi-infinite rod.}
\label{fig:varilla}
\end{figure}

\begin{figure}
\includegraphics[width=8.3cm]{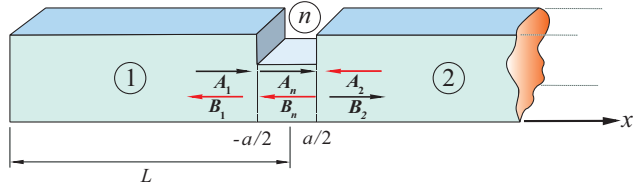}
\caption{Waves in the different regions of the semi-infinite rod. The width and
depth of the notch are $a$ and $h$, respectively. The length of the cavity is
$L-a/2$.}
\label{fig:ondas}
\end{figure}

In order to study the resonances of an elastic one-dimensional cavity, lets
consider a semi-infinite elastic rod with square cross section of side $W$. As it is shown in
Fig.~\ref{fig:varilla}, the cavity is formed by a rectangular notch of width $a$
and depth $h$ which have been machined at a distance $L$ from the free-end of
the rod. To first order, the torsional wave equation gives a correct description
of the scattering of the waves in all regions: inside the cavity, at the
notch, and outside the cavity. This model is an analogue of a quantum mechanical
delta potential situated at a certain distance from an impenetrable {\em soft}
wall (Neumann boundary conditions). The solution of the torsional wave equation, in terms of the wave amplitudes in the different regions of the rod (see Fig.~\ref{fig:ondas}), can be written as 
\begin{equation}
\psi(x) = \left\{ \begin{array}{ll}
A_{1} e^{ikx} + B_{1} e^{-ikx} & \textrm{for } -L \leq x \leq -a/2\\
A_{n} e^{ik_{n}x} + B_{n} e^{-ik_{n}x} & \textrm{for } -a/2 \leq x \leq a/2\\
A_{2} e^{-ikx} + B_{2} e^{ikx} & \textrm{for } x \geq a/2
\end{array} \right. , 
\end{equation}
where the wave number $k_j$, in the corresponding region of the beam, is
given by~\cite{Graff}
\begin{equation}
\label{eq:wavenumber}
k_j = \frac{2\pi}{c_j} f, 
\end{equation}
with $f$ the frequency and $c_j$ the velocity of the waves in the respective
region. This velocity is related to the shear
modulus $G$ and the density $\rho$, through
\begin{equation}
c_j = \sqrt{\frac{G}{\rho} \frac{\alpha_j}{I_j}},
\end{equation}
where $I_j$ is the polar momentum of inertia and $\alpha_j$ is given by the
Navier series in the corresponding region with rectangular cross section whose
base is $W_j$ and height $h_j$ [for $x\in(-L,-a/2)$, $h_1=W_1=W$, while
$h_n=W-h$ and $W_n=W$ for $x\in(-a/2,a/2)$]; that
is, 
\begin{equation}
\alpha_j = \frac{256}{\pi^6} \sum_{m=0}^{\infty} \sum_{p=0}^{\infty}
\frac{1}{(2m + 1)^2 (2p + 1)^2} \frac{h_j W_j}{(\frac{2m + 1}{h_j})^2 +
(\frac{2p + 1}{W_j})^2}\,. 
\end{equation}

Since the one-dimensional cavity has a free-end at $x=-L$, we impose the
condition that $d\psi(x)/dx$ vanishes at $x=-L$. We will see below that this
boundary condition gives an appropriate description of the experiment. The
continuity of $\psi(x)$, as well as of the torsional momentum between the
different regions, allows us to obtain the scattering matrix associated to the
system, namely 
\begin{equation}
S = r_{n} + t_{n} \frac{1}{1 - r_{n} e^{2ik(L-a/2)}} e^{2ik(L-a/2)} t_{n},
\label{eq:Smatrix}
\end{equation}
where $r_n$ and $t_n$ are the reflection and transmission amplitudes through the
notch, given by
\begin{equation}
r_n = -\frac{e^{ik_na} - e^{-ik_na}}{\frac{\lambda - 1}{\lambda + 1} e^{ik_na} -
\frac{\lambda + 1}{\lambda - 1} e^{-ik_na}}
\end{equation}
and
\begin{equation}
t_n = \frac{\frac{\lambda - 1}{\lambda + 1} - \frac{\lambda + 1}{\lambda - 1}}
{\frac{\lambda - 1}{\lambda + 1} e^{ik_na} - \frac{\lambda + 1}{\lambda - 1}
e^{-ik_na}} ,
\end{equation}
where $\lambda=\alpha_2/\alpha_1$. 

Notice that the scattering matrix depends on the frequency through the wave
numbers [see Eq.~(\ref{eq:wavenumber})]. With ideal conditions $S(f)$ is an
unitary matrix, such that in the $1\times 1$ case it becomes a complex number of
unit modulus. In a real situation when $S(f)$ is measured in arbitrary
units, as it is the case of elastic systems, its modulus is not unit; that is,
$S(f)=\sqrt{R_0}\,e^{i\theta(f)}$. Therefore, the movement of $S(f)$ as $f$ is 
varied, describes a circle of radius $\sqrt{R_0}$ in the Argand plane; but 
it does not visit the circle with the same probability; instead $S(f)$ is 
distributed according to the non-unitary Poisson's
kernel~\cite{Martinez-ArguelloMendez-SanchezMartinez-Mares}
\begin{equation}
\label{eq:PoissonNonUnitary}
p(\theta) = 
\frac{1}{2\pi} \frac{R_0 - \left| \overline{S} \right|^2}
{\left| S - \overline{S} \right|^2}, 
\end{equation}
where $\overline{S}$ is the average of $S(f)$ in frequency which, together with
$R_0$, is obtained from the experiment. Of course, when $R_0=1$,
Eq.~(\ref{eq:PoissonNonUnitary}) reduces to the ordinary Poisson's
kernel~\cite{MelloSeligmanPereyra}.

\begin{figure}
\includegraphics[width=0.5\columnwidth]{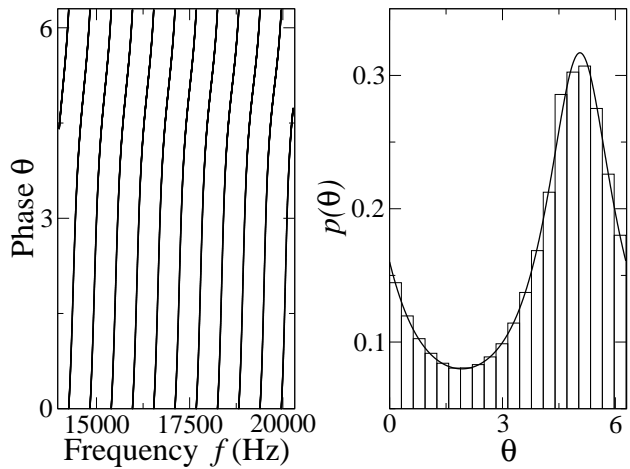}
\caption{The phase $\theta$ of the $S$-matrix in the frequency range between
$14$~kHz and $20$~kHz (left panel). The distribution of $\theta$, for the same 
numerical data (histogram), agrees with Poisson's kernel, Eq~(\ref{eq:PoissonNonUnitary})
(continuous line in the right panel).}
\label{fig:MS}
\end{figure}

For the elastic cavity modeled by Eq.~(\ref{eq:Smatrix}), $R_0=1$ and $\theta$
depends on the frequency but also on the parameters of the rod which are
fixed. As an example, we consider a cavity formed on an aluminum rod of square
cross section of 25.4~mm of side with a notch of 18.0~mm depth. The physical parameters
of the aluminum alloy 6061-T6, that we use in the experiment,
are $G=26$~GPa and $\rho=2.7$~g/cm$^3$, such that
$\sqrt{G/\rho}=3103.2$~m/s. The numerical resonances of this
cavity, eleven in total, are shown in Fig.~\ref{fig:MS}, where $\theta(f)$ is plotted as a
function of $f$ for the frequency range between $14$~kHz and $20$~kHz. Also, in Fig.~\ref{fig:MS} we show the distribution
of $\theta$ in this frequency range, which has been obtained numerically and
compared with Poisson's kernel given by Eq.~(\ref{eq:PoissonNonUnitary}) for
$R_0=1$. The agreement is almost perfect.

\section{The Experimental Setup}
\label{sec:ExperimentalSetup}

\begin{figure}
\includegraphics[width=0.5\columnwidth]{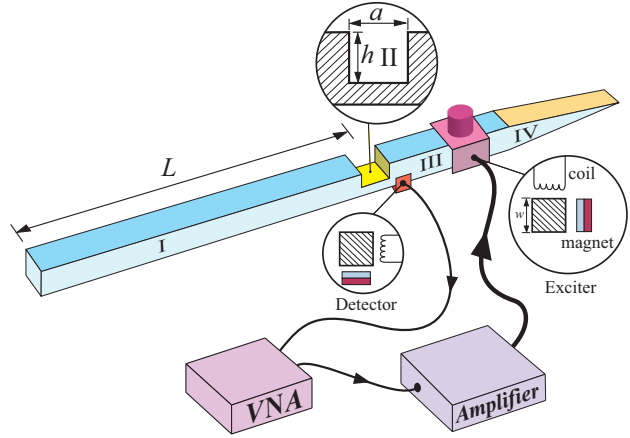}
\caption{In the system $L=2.5$~m, the depth and width of the notch are
$h=0.018$~m and $a=0.0009$~m, respectively. The beam has a total length of
$3.6$~m and a square cross-section of $0.0254$~m of side. The wedge has a length
of $0.40$~m and it is covered by a polymeric foam. The elastic beam is suspended
by two nylon strings (not shown). The connection of the equipment is shown at
the bottom.}
\label{fig:ExperimentalSetup}
\end{figure}

The theoretical model described in the previous section can be studied
experimentally in the corresponding elastic system. As it is seen in
Fig.~\ref{fig:ExperimentalSetup}, we use an aluminum rod with square cross
section. This rod is divided in four regions: Region I is the
quasi-one-dimensional cavity of length $L-a/2$, where $a$ is the width of a
notch of depth $h$, that defines Region~II and simulates the quantum delta
potential. Regions III and IV mimic the semi-infinite one-dimensional space at the
right of the cavity; the vibrations are trapped in a wedge in Region~IV
which acts as a passive attenuation system, together with a polymeric foam that cover it.
This scheme minimizes the reflection at the right-end of the rod and
consequently the normal modes of the complete system are diminished; it allows to
measure the resonances of the cavity formed by the left free-end of the rod and
the notch.

The elastic system is subject to a torsional elastic excitation {\em via} an
electromagnetic acoustic transducer (EMAT) disposed in torsional wave
configuration~\cite{MoriFloresGutierrez}. The exciter generates a sinusoidal
torque at Region III of the elastic system. The excitation of the wave is
produced by an oscillating magnetic field of frequency $f$ generated by an
AC current $I(t)$ on the EMAT's coil, at the same frequency. When a
paramagnetic metal is close to the EMAT magnetic field, as Faraday's induction
law establishes, eddy currents are produced inside the metal. These currents
experience the Lorentz force due to the permanent magnetic field of the EMAT's
magnet. In consequence the metal rotates locally in both directions at the frequency
$f$. 

The response is detected by a second EMAT located outside the cavity, as shown
in Fig.~\ref{fig:ExperimentalSetup}. As a detector, the EMAT works in the
following way: when a rotating metal is located near the field of a permanent
magnet, some loops in the metal will have a non-vanishing variable magnetic flux
that will produce eddy currents. These currents will generate an oscillating
magnetic field that will be measured by the EMAT's coil. In this way, the EMAT
detector measures the torsional acceleration of diamagnetic metal~\cite{refEMAT}.

To produce the torsional vibration in the elastic system we use a vector network
analyzer (VNA, Anritsu MS-4630B). The VNA produces a sinusoidal signal of
frequency $f$, which is amplified by a Cerwin-Vega CV-900 amplifier. The
amplified signal is sent to the EMAT exciter. The torsional acceleration
measured by the EMAT detector is recorded directly by the VNA for its analysis (see
Fig.~\ref{fig:ExperimentalSetup}).

\begin{figure}
\includegraphics[width=0.5\columnwidth]{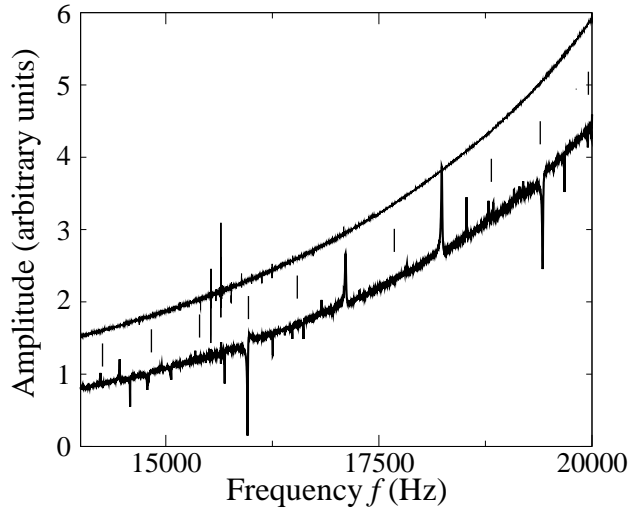}
\caption{Spectrum measured with the detector located just outside the cavity  after the
notch (thick line). The thin line corresponds to the base
line measured without the magnet of the EMAT detector. The vertical lines
correspond to the theoretical predictions obtained from Fig.~\ref{fig:MS} (see also
Table~\ref{table:comparacion} below).}
\label{fig:Spectrum}
\end{figure}

\section{Comparison between theory and Experiment}
\label{sec:Comparison}

The spectrum of a typical measurement is given in Fig.~\ref{fig:Spectrum}, which
shows the observed resonances (thick line). The thin line corresponds to a
measurement in which the magnet of the EMAT has been removed; this signal
is the impedance curve of the coil and the lines that appear over it are
due to radio stations and they must be disregarded. One can observe, also in
Fig.~\ref{fig:Spectrum}, that the experimental resonances of the cavity are in
very good agreement with the theoretical predictions (vertical marks) of
Fig~\ref{fig:MS}. The comparison between the numerical values of the resonances
of torsional waves and the predicted ones, is given in
Table~\ref{table:comparacion}, where errors less than 0.1\% are observed.
Although some resonances do not appear, they become visible when the location
of the EMAT exciter is changed. The remaining resonances belong to other types of
vibrations (compressional or flexural).  

\begin{table}
\centering 
\begin{tabular}{|c|c|c|c|} 
\hline\hline 
Resonance & Theory (Hz)  & Experiment (Hz) \\ 	
\hline 
1  & 14256  & ------ \\ 
2  & 14826  & 14819 \\ 
3  & 15396  & ------ \\ 
4  & 15966  & 15960 \\ 
5  & 16537  & 16528\\ 
6  & 17107  & 17097 \\
7  & 17677  & ------ \\
8  & 18247  & 18237 \\
9  & 18818  & 18800 \\
10 & 19388  & 19400 \\
11 & 19958  & 19949 \\ 
\hline 
\end{tabular}
\caption{Torsional resonances of a cavity of length $L=2.5$~m formed by one free
end and a notch of width $a=0.0009$~m and depth $h=0.018$~m. The agreement
between the theory and experiment is excellent.}
\label{table:comparacion} 
\end{table}

\begin{figure}
\includegraphics[width=0.6\columnwidth]{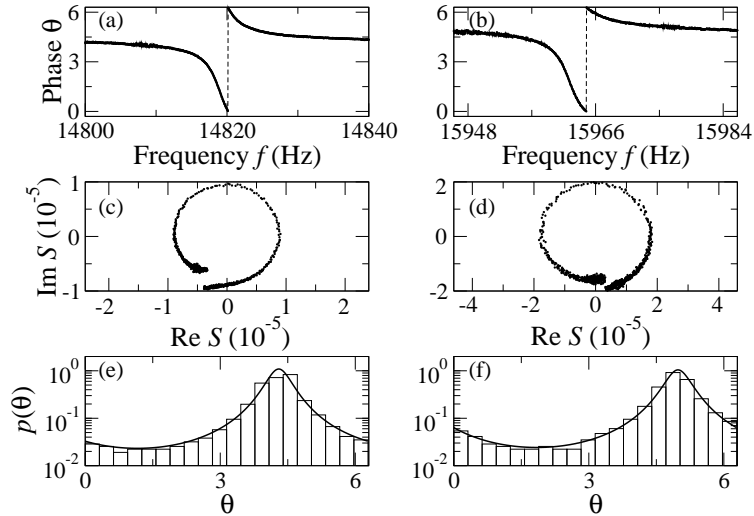}
\caption{Phase of the $S$-matrix as a function of frequency $f$ (a) for the
resonance at 14819~Hz and (b) for the resonance at 15960~Hz. The movement of the
$S$-matrix as a function of frequency $f$ in the Argand diagram for the same
resonances are given in the panels (c) and (d), respectively. The distributions
of the phase in panels (e) and (f) show an excellent agreement with the
non-unitary Poisson kernel (continuous line). The resonances are observed from
the center of the circles (for explanation see the text).}
\label{fig:ExperimentalResonances}
\end{figure}

Within the frequency range measured, between 14 kHz and 20 kHz, there are eleven
resonances. Due to the impedance of the EMAT's coil, the scattering matrix $S$
describes a circle in the Argand diagram, but displaced from the origin (not
shown here). In Fig.~\ref{fig:ExperimentalResonances} we show two of the
resonances, (a) 14819~Hz and (b) 15960~Hz (see Table~\ref{table:comparacion}),
as they are seen from the center of their corresponding circles. In panels (c)
and (d) of Fig.~\ref{fig:ExperimentalResonances} we observe the circles
whose radii are not the unit. As we previously explain, this
is due to the arbitrary units of measurement in the amplitude. In panels (e) and
(f) we compare the histograms of their phases with the non-unitary Poisson
kernel given by Eq.~(\ref{eq:PoissonNonUnitary}), where the values of the
average $\overline{S}$ of the $S$-matrix and their radii, taken as averaged
quantities, have been extracted from the corresponding experimental data. As can
be seen, the agreement is excellent still for the resonance at $f=14819$~Hz that
shows the worst agreement.

\section{Conclusions}
\label{sec:Conclusions}

We have studied the scattering of torsional waves in a quasi-one-dimensional
elastic system. This system consists of a beam with a notch between a free-end
and a passive vibration attenuation system that
simulates the incoming and outgoing channels at one end of the rod. Theoretically, we obtained the
$1\times 1$ scattering matrix from the solution of the torsional wave equation;
the numerical predictions helped to select the torsional resonances, among
many other vibrational modes that were detected in the experiment. We
also verified that the experimental distribution of the phase of the scattering
matrix is described by Poisson's kernel, which is a very important theoretical
result in scattering of waves by open systems. Despite that the rod is finite we have
confirmed that we opened the system from one side, forming in this way a
quasi-one-dimensional open cavity. 

\section{Acknowledgements}

This work was supported by DGAPA-UNAM under project PAPIIT IN111311. MCS and
AMMA thank financial support from CONACyT. MMM is grateful with the Sistema
Nacional de Investigadores and MA Torres-Segura for her encouragement.

\end{document}